\documentclass[manuscript,screen]{acmart}

\AtBeginDocument{%
  }

\setcopyright{acmlicensed}
\copyrightyear{2018}
\acmYear{2018}
\acmDOI{XXXXXXX.XXXXXXX}
\acmConference[Conference acronym 'XX]{Make sure to enter the correct
  conference title from your rights confirmation email}{June 03--05,
  2018}{Woodstock, NY}
\acmISBN{978-1-4503-XXXX-X/2018/06}

\usepackage{siunitx}

\begin{document}

%%
%% The "title" command has an optional parameter,
%% allowing the author to define a "short title" to be used in page headers.
\title[Meander \textit{NFC} and picoRing \textit{NFC}]{Body-scale NFC for wearables: human-centric body-scale NFC networking for ultra-low-power wearable devices (Demo of UTokyo Kawahara Lab)}

%%
%% The "author" command and its associated commands are used to define
%% the authors and their affiliations.
%% Of note is the shared affiliation of the first two authors, and the
%% "authornote" and "authornotemark" commands
%% used to denote shared contribution to the research.

\author{Hideaki Yamamoto*}
\affiliation{%
  \institution{The University of Tokyo}
  \city{Tokyo}
  \country{Japan}
  }
\email{yamamoto-hideaki@akg.t.u-tokyo.ac.jp}
\orcid{0009-0002-8977-8371}

\author{Yifan Li}
\authornote{Two authors contributed equally to this research.}
\affiliation{%
  \institution{The University of Tokyo}
  \city{Tokyo}
  \country{Japan}
  }
\email{yifan217@akg.t.u-tokyo.ac.jp}
\orcid{0009-0005-9261-6391}

\author{Wakako Yukita}
\affiliation{%
  \institution{The University of Tokyo}
  \city{Tokyo}
  \country{Japan}
  }
\email{yukita@bhe.t.u-tokyo.ac.jp}
\orcid{0009-0003-4043-2318}

\author{Tomoyuki Yokota}
\affiliation{%
  \institution{The University of Tokyo}
  \city{Tokyo}
  \country{Japan}
  }
\email{yokota@ntech.t.u-tokyo.ac.jp}
\orcid{0000-0003-1546-8864}

\author{Takao Someya}
\affiliation{%
  \institution{The University of Tokyo}
  \city{Tokyo}
  \country{Japan}
  }
\email{someya@ee.t.u-tokyo.ac.jp}
\orcid{0000-0003-3051-1138}

\author{Ryo Takahashi$^{\dagger}$}
\affiliation{%
  \institution{The University of Tokyo}
  \city{Tokyo}
  \country{Japan}
  }
\email{takahashi@akg.t.u-tokyo.ac.jp}
\orcid{0000-0001-5045-341X}

\author{Yoshihiro Kawahara}
\authornote{Correspondence to Y. Kawahara and R. Takahashi}
\affiliation{%
  \institution{The University of Tokyo}
  \city{Tokyo}
  \country{Japan}
  }
\email{kawahara@akg.t.u-tokyo.ac.jp}
\orcid{0000-0002-0310-2577}

\renewcommand{\shortauthors}{H. Yamamoto, Y. Li, W. Yukita, T. Yokota, T. Someya, R. Takahashi, Y. Kawahara}

\begin{abstract}
Near Field Communication (NFC) is a promising technology for ultra-low-power wearables, yet its short communication range limits its use to narrow-area, point-to-point interactions.
We propose a body-scale NFC networking system that extends NFC coverage around the body, enabling surface-to-multipoint communication with distributed NFC sensor tags.
This demonstration introduces two key technologies: Meander NFC and picoRing NFC.
First, Meander NFC expands a clothing-based NFC networking area up to body scale while enabling a stable readout of small NFC tags occupying \qty{1}{\%} of the coverage area.
Meander NFC uses a meander coil which creates a spatially confined inductive field along the textile surface, ensuring robust coupling with small tags while preventing undesired electromagnetic body coupling.
Second, picoRing NFC solves the weak inductive coupling caused by distance and size mismatches.
By leveraging middle-range NFC and coil optimization, picoRing NFC extends the communication range to connect these disparate nodes between the ring and wristband.

\end{abstract}

%%
%% The code below is generated by the tool at http://dl.acm.org/ccs.cfm.
%% Please copy and paste the code instead of the example below.
%%


%% A "teaser" image appears between the author and affiliation
%% information and the body of the document, and typically spans the
%% page.
\begin{teaserfigure}
  \includegraphics[width=1\textwidth]{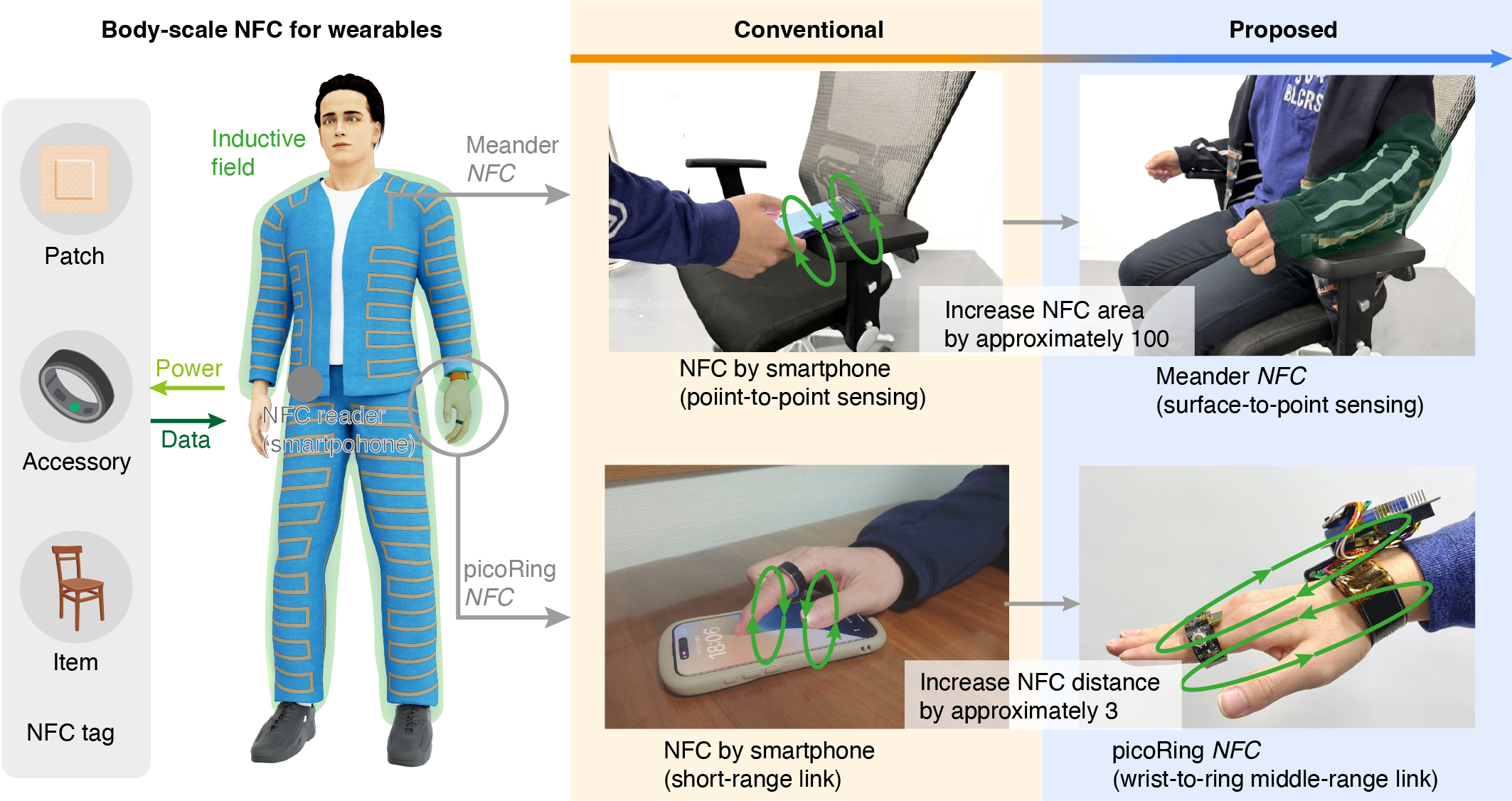}
  \caption{Overview of body-scale NFC for wearables.}
  \Description{}
  \label{fig:teaser}
\end{teaserfigure}

\received{20 February 2007}
\received[revised]{12 March 2009}
\received[accepted]{5 June 2009}

%%
%% This command processes the author and affiliation and title
%% information and builds the first part of the formatted document.
\maketitle

\section{Introduction}

The evolution of wearable computing is moving toward a distributed ecosystem where multiple wearable devices are spread across the human body to capture diverse physiological and interaction data~\cite{kim_epidermal_2011,yu_intelligent_2026}. 
To realize this vision, recent advances in wearable devices have increasingly demanded ultra-low-power or battery-free architectures to ensure long-term, unobtrusive monitoring and interaction~\cite{zheng_unobtrusive_2014}.
While the sensing and computation components of a wearable system can operate at ultra-low-power levels below microwatts or even nanowatts, data communication remains a significant energy bottleneck. 
Conventional wireless technologies, such as Bluetooth and Wi-Fi, typically consume tens to hundreds of milliwatts at maximum, which is orders of magnitude higher than the power budget of ultra-low-power wearables~\cite{wagih_microwave-enabled_2023}. 
Near-field communication (NFC), based on near-field backscatter technology, has emerged as a promising alternative for ultra-low-power wearables due to its low power consumption, typically below \qty{1}{\mW}~\cite{zhao_nfc-wisp_2015}. 
However, NFC's inherent limitation of a short communication range (typically a few centimeters) restricts its applicability to short-range, point-to-point interactions, making it challenging to establish a body-scale network of wearable devices. 
For instance, battery-free epidermal patch sensors require the user to manually bring a smartphone equipped with an NFC reader within a few centimeters to log data, which imposes temporal constraints on the NFC networks~\cite{zou_nfcrfid-enabled_2025,oh_battery-free_2021}.
Furthermore, NFC-enabled smart rings and wristbands cannot exchange data with each other if they are located on different limbs, creating spatial constraints for wearable usage~\cite{takahashi_picoring_2024,li_ultra-low-power_2025}.

To address these limitations, we propose a body-scale NFC networking system named \textit{body-scale NFC for wearables} that extends NFC temporal and spatial coverage across the entire human body, facilitating scalable and energy-efficient communication with distributed sensor tags.
This demonstration introduces two key technologies: Meander \textit{NFC} and picoRing \text{NFC} (see \autoref{fig:teaser}).
Meander \textit{NFC} overcomes the temporal constraints of traditional NFC readers by integrating a body-scale NFC coil into clothing. 
Unlike previous coils that radiate magnetic fields spatially, Meander \textit{NFC} employs a meander coil that confines the strong magnetic field near the skin. 
This surface-like wireless space ensures stable readout even when small sensor tags occupy only \qty{1}{\percent} of the large-area coil, enabling always-on monitoring without requiring the user to manually position a reader.
Our demonstration uses lightweight and low-loss copper foils, enabling comfortable long-term wearability without compromising NFC performance, unlike previous lossy conductive yarns~\cite{takahashi_twin_2021,takahashi_full-body_2025} or low-loss but heavy liquid metal tubes~\cite{takahashi_meander_2022,sato_friction_2025,takahashi_full-body_2026}.
Second, picoRing \textit{NFC} uses a middle-range NFC technique to demonstrate reliable NFC communication between a ring and a wristband, even under weak inductive coupling over \qty{10}{\cm} distance.
Unlike previous picoRing based on passive inductive telemetry~\cite{takahashi_picoring_2024,li_ultra-low-power_2025} --- one-way, slow, frequency-sweeping readout of the wristband coil to the ring $LC$ tank---, picoRing \textit{NFC} enables two-way, fast, time-division communication at kilobytes of data.
The fast, two-way communication allows for interactive control: a gesture detected by the ring can be used to adjust parameters on the wristband display (e.g., volume control, data visualization settings) or trigger actions in connected applications on AR glasses.

\section{Related work}

Our work builds upon a rich history of research in Near-Field Communication (NFC) protocols and their applications within Human-Computer Interaction (HCI). 
NFC emerged as a subset of radio-wave frequency identification (RFID), initially focused on short-range identification and contactless payment.
The system operates through the inductive coupling between two coils: the reader (initiator) and the tag (target). 
The reader generates an inductive field at a carrier frequency of \qty{13.56}{\mega\hertz}, which induces a current in the tag's coil. 
This process provides sufficient wireless power to activate the internal microchip of the tag.
Once activated, the tag transmits data back to the reader by varying its electrical load, known as backscatter or load modulation. 
By altering the reflected signal strength, the tag sends its data with low power consumption, making it ideal for battery-free wearable applications.
The core NFC standards (ISO/IEC 14443 Type A, B, and F) have traditionally been limited to proximity-based interactions. 
However, recent processes have expanded the protocol's utility: narrow-band NFC (ISO/IEC 15693 Type V) enables middle-range communication up to \qtyrange{5}{10}{\cm}~\cite{zhao_nfc_2020}, while the NFC Wireless Charging specification now allows for reliable power transfer of up to \qty{2.2}{\W} (ST25R300, STMicroelectronics). 
This power capability is particularly valuable for the advanced and power-consuming hundreds of \si{\mW}-class wearable devices such as smartwatches and smartglasses. 
While early HCI research utilized these protocols for tangible interfaces and simplified device pairing~\cite{hardy_touch_2008,fei_peripheral_2013}, recent research have integrated NFC tags into jewelry~\cite{lee_nfcstack_2022,zhao_nfc-wisp_2015}, clothing~\cite{ye_body-centric_2022,lin_wireless_2020,noda_wearable_2019,hajiaghajani_textile-integrated_2021,hajiaghajani_amphibious_2023}, furniture~\cite{wang_locating_2023,oh_battery-free_2021}, capsule~\cite{zhang_nfcapsule_2023}, patch~\cite{cai_osseosurface_2021,shao_room-temperature_2022,yamagishi_flexible_2024}, and gadgets~\cite{ma_nfinger_2026,yamagami_morphkeys_2025,li_nfcgest_2025,dai_magcode_2023,saito_japanese_2020,an_one_2021,liang_nfcsense_2021,villar_project_2018} for gesture recognition and physiological monitoring~\cite{zou_nfcrfid-enabled_2025}.

\section{Design}

\begin{figure}[t!]
  \centering
  \includegraphics[width=\columnwidth]{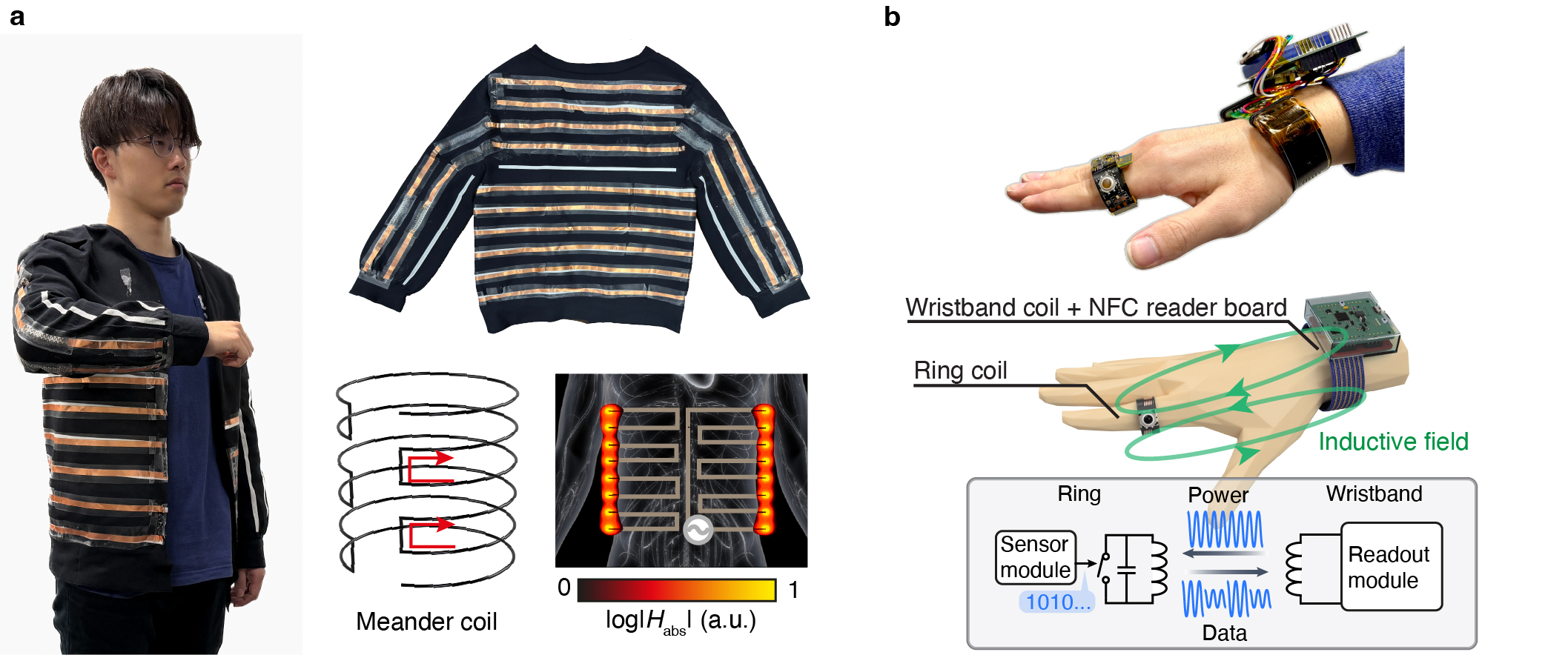}
  \caption{Design schematics of Meander (a-c) \textit{NFC} and (d-f) picoRing \textit{NFC}.}
  \label{fig:design}
  \Description{}
\end{figure}

\subsection{Meander \textit{NFC}}

Meander \textit{NFC} is designed to provide continuous, wireless power and data networking across the large-area, dynamic surfaces of the human body. 
The core of Meander \textit{NFC} consists of a meandered textile coil fabricated by copper foils (see \autoref{fig:design}a). 
The meander pattern, which changes its winding direction for each turn, is designed to confine the magnetic field close to the skin --- \textit{two-dimensional localized inductive field} ---, where wearable devices are typically placed.
Therefore, unlike conventional spiral or helical loop coils, which spatially radiate the inductive field, the meander coil allows for energy-efficient power and data transfer to NFC tags even near the dielectric human body.
In addition, even when the textile coil is deformed due to dynamic movements of the human body, Meander NFC ensures a stable NFC link with the tag.
This is because the localized effect can function as long as the local wire spacing of the meander coil remains almost constant.

Our prototype uses lightweight and low-loss copper foils with \qty{8}{\um} thickness, enabling comfortable wearability without compromising NFC performance. 
This approach overcomes the limitations of previous methods, such as lossy conductive yarns~\cite{takahashi_twin_2021,takahashi_full-body_2025} or conductive but heavy liquid metal tubes~\cite{takahashi_meander_2022,sato_friction_2025,takahashi_full-body_2026}.
We implement two meander coils on the garment: one covering the abdominal area and another extending across the arms and back.
The wiring patterns are cut using a mechanical plotter or a UV laser. 
To ensure stretchability at joints such as the elbows and shoulders, we employ a double-track serpentine wiring, providing mechanical durability for the user's motion.
In total, the inductance, $Q$-factor, and power transfer efficiency of the abdominal/arm-sleeve meadner coils with customized NFC sensor tag (STEVAL-SMARTAG1, STMicroelectronics) shows \qty{3.4}{\uH}/\qty{2.7}{\uH}, \num{95}/\num{53}, and \qty{41}{\%}/\qty{30}{\%}, respectively. 
The average power consumption of the NFC reader board and NFC sensor tag is $\qty{515}{\mW} (= \qty{5.0}{\V} \times \qty{103}{\mA})$ and $\qty{845}{\uW} (= \qty{3.3}{\V} \times \qty{250}{\uA})$, respectively.

\begin{figure}[t!]
  \centering
  \includegraphics[width=\columnwidth]{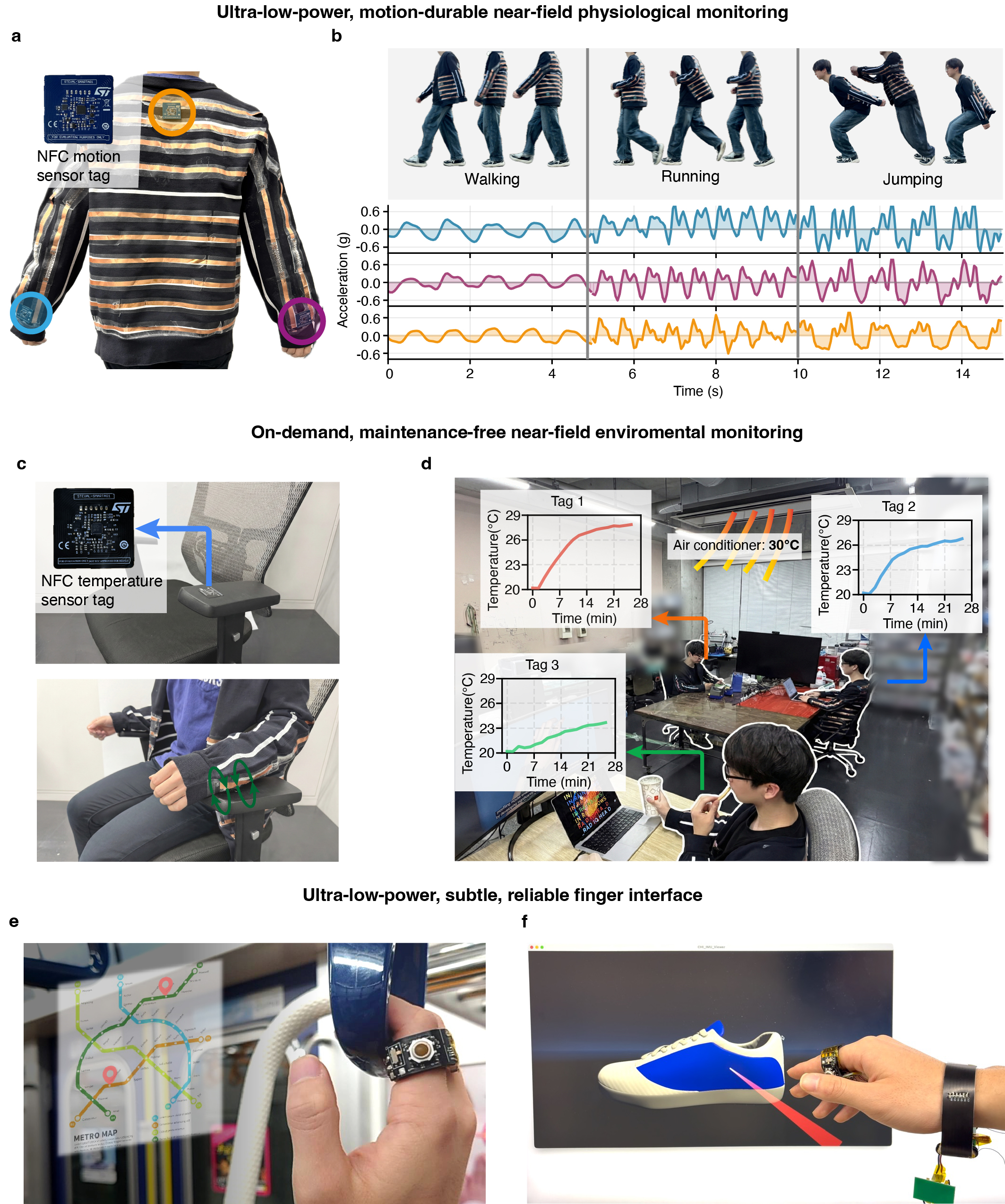}
  \caption{Application examples of Meander \textit{NFC} and picoRing \textit{NFC}.}
  \label{fig:app}
  \Description{}
\end{figure}

\subsection{picoRing \textit{NFC}}

The picoRing \textit{NFC} is a compact, finger-worn device designed for reliable, subtle finger input with ultra-low power (see \autoref{fig:design}b). 
Building upon the original picoRing~\cite{takahashi_picoring_2024,li_ultra-low-power_2025}, picoRing \textit{NFC} integrates a low-power, high-speed 3-axis accelerometer and an NFC communication module within a lightweight ring form factor. 
This turns the finger into a micro finger interaction hub, where the user can perform precise tactile tasks (e.g., scrolling or clicking) while the ring concurrently captures hand gestures.
To link the wristband with the small ring through NFC protocols, picoRing \textit{NFC} combines the NFC Type V with an angled coil design.
This configuration enables to maximize coil size within the compact ring and enhance the inductive coupling between the ring and wristband.
Furthermore, by utilizing NFC's optimized sleep modes, picoRing \textit{NFC} maintains the long-lasting battery life required for continuous daily use.
In total, the $Q$-factor and power consumption (active and sleep modes) of the ring/wristband coil shows \num{23.9}/\num{15.9}, and $\qty{2.02}{\mW} (= \qty{1.8}{\V} \times \qty{1.12}{\mA})$/$\qty{705}{\mW} (= \qty{5.0}{\V} \times \qty{141}{\mA})$ (active) and $\qty{371}{\uW} (= \qty{1.8}{\V} \times \qty{206}{\uA})$/$\qty{83.5}{\mW} (= \qty{5.0}{\V} \times \qty{16.7}{\mA})$ (sleep), respectively.

\section{Demonstration}

Meander \textit{NFC} introduces a flexible approach to wearable sensing technology by providing a body-scale and motion-robust wireless sensing surface that works regardless of where the NFC sensor tags are placed.
While previous research, such as e-textiles composed of multiple sparse small coils~\cite{lin_wireless_2020,lin_digitally-embroidered_2022}, is effectively available for specific applications, these systems are rigid in their design.
If the application changes—for example, moving from tracking users' posture to monitoring surrounding environmental signals, users need to be redesigned to match the new sensor locations.
In contrast, Meander \textit{NFC} turns the garment itself into a universal near-field sensing e-textile platform. 
Users can freely place tags without changing the wiring of the clothes. 
This eliminates the need for application-specific redesigns of the near-field e-textiles.
Meander \textit{NFC} enables high-fidelity motion tracking by simply placing the NFC sensor tags on the arms or waist (see \autoref{fig:app}a). 
Even when the textile meander coils are deformed due to body movement, Meander \textit{NFC} enables reliable power and data transfer in any posture (see \autoref{fig:app}b).
This is because the generated inductive field remains stably confined near the surface of the clothing as long as the local spacing of the wire pattern does not change drastically.
In addition, the body-scale coverage of Meander \textit{NFC} allows the clothing to interact with the surrounding environment and objects around it. 
By placing the temperature NFC tags on a chair, Meander \textit{NFC} automatically records the local environment the moment the user sits down, because the NFC connection is guaranteed at arbitrary points where the clothes touch the furniture (see \autoref{fig:app}c).
This turns the user into a mobile sensing platform, capable of monitoring room-wide air conditioning levels through simple, everyday actions (see \autoref{fig:app}d).

picoRing \textit{NFC} shows ultra-low-power ring input interfaces for continuous subtle finger interaction with wearable digital content provided by smartwatch ans smartglasses. 
While the original picoRing was designed for cursor control using a microtrackball~\cite{li_ultra-low-power_2025}, the new picoRing \textit{NFC} integrates a motion sensor in addition to its tactile input. 
This allows the ring to reliably capture both precise finger movements and the broader dynamic 3D motion of the hand.
By moving beyond simple scrolling and clicking, picoRing \textit{NFC} enables a dual-mode interaction style. 
Users can perform traditional digital tasks with the trackball while simultaneously tracking complex hand gestures or physical activities through the motion sensor (see \autoref{fig:app}e-f). 
%This compact form factor turns a single finger into a powerful sensing hub, providing high-fidelity data in a device that is small enough to be worn comfortably throughout the day.

\section{Conclusion}
\label{sec:conclusion}

In this demonstration, we present Meander \textit{NFC} and picoRing \textit{NFC}.
Unlike previous e-textile networks that require application-specific layouts, Meander \textit{NFC} provides a body-scale wireless power and data networking that maintains stable wireless connectivity even during movements like running or jumping. 
This allows users and researchers to deploy sensors wherever they are needed without redesigning the garment.
Complementing this body-scale system, picoRing \textit{NFC} supports both long-lasting battery life required for daily wear and fast data transfer like motion data, in addition to trackball inputs.
This enriches the finger-tip interface, allowing it to capture rapid 3D gestures and subtle tactile commands simultaneously.
Therefore, our demonstration showcases how this combination of wide-area sensing and rich, energy-efficient finger-tip control can support a broad range of applications, from personalized health monitoring to subtle finger interaction in daily life.

\begin{acks}
This work was mainly supported by JST CRONOS JPMJAX21K9, JST PRESTO JPMJPR2515, JSPS KAKEN 22K21343, JST ASPIRE JPMJAP2401, and the Asahi Glass Foundation.
The authors used Google Gemini for linguistic refinement and editing of this manuscript.
\end{acks}

%%
%% The next two lines define the bibliography style to be used, and
%% the bibliography file.
\bibliographystyle{ACM-Reference-Format}
\bibliography{references}

\end{document}